\documentclass[aip,jcp,reprint,onecolumn,12pt,amsmath,amssymb,amsfonts,floatfix,letterpaper]{revtex4-1}

\usepackage{graphicx}
\usepackage{braket}
\usepackage{tablefootnote}
\usepackage{color}
\usepackage{enumitem}

\begin{document}

\title{General-order open-shell coupled-cluster method with partial-spin adaptation II: further formulations, simplifications, implementations, and numerical results}
\author{Cong Wang}
\email{congwang.webmail@gmail.com}
\affiliation{PO Box 26 Okemos, MI, 48805 (USA)}
\affiliation{The Pennsylvania State University; 401A Chemistry Building; University Park, PA 16802 (USA)}

\begin{abstract}

This is a continuation of the previous work (arXiv:2403.10128). Additional aspects such as linear combinations of projections and hash-table canonicalizations are described.  Implementations of the general-order partial-spin adaptation (PSA) coupled-cluster (CC) method are outlined.  Numerical results are reported.

\end{abstract}

\maketitle

\section{Introduction}

Adapting spatial orbitals in open-shell CC methods  \cite{bartlett1995coupled, christiansen2006coupled, Bartlett:07,crawford2007introduction,Bartlett:09, bartlett2012coupled,     Lyakh:12,bartlett2024perspective, Helgaker:00book,Harris:99,Olsen:00,Hirata:00, hirata2003tensor, Kallay:00,kallay2001higher, auer2006automatic, parkhill2010sparse,engels2011fully,evangelista2022automatic,Stanton:14,Stanton:15} can reduce the prefactors of computational costs comparing to the spin-orbital approaches \cite{Bartlett:09} at least in principle, since the dimension the vector space of interest from the spatial indices is smaller than the spin-orbital indices for a given basis set \cite{Helgaker:00book}. Different from the closed-shell approaches \cite{scuseria1987closed, scuseria1988efficient, lee1988efficient, koch1990coupled,hampel1992comparison,Helgaker:00book,Bartlett:09,Stanton:14,matthews2014non,Stanton:15, wang2018simple, matthews2019diagrams}, the contractions between active orbitals in open-shell systems complicate the formulations and caused open-shell spatial-orbital CC methods remain a challenge in quantum chemistry \cite{lindgren1978coupled, nakatsuji1977cluster,  nakatsuji1978cluster_a, nakatsuji1978cluster, mukherjee1979hierarchy,  nakatsuji1979cluster,nakatsuji1983cluster,haque1984application, mukherjee1989use, Janssen:91,Knowles:93,jayatilaka1993open,Li:94,neogrady1994spin, Jeziorski:95a,Neogrady:95,nooijen1996many, nooijen1996general,urban1997spin, Szalay:97, Szalay:00,Li:98,Jeziorski:99,nooijen2001towards, grotendorst2000modern,knowles2000erratum, Helgaker:00book,Heckert:06,Datta:08, datta2011state, Datta:13,datta2014analytic, gunasekera2024multireference, datta2015communication,datta2019accurate, herrmann2020generation,herrmann2022analysis,herrmann2022correctly,folkestad2023entanglement}.   Accuracy beyond the chemical accuracy (1 kcal mol$^{-1}$) can be important in predicting reaction selectivities \cite{krieger2021impact,laplaza2024overcoming}. Higher-order CC methods (beyond triple excitations in the cluster operator) \cite{martin1999towards,tajti2004heat,yang2014ab,karton2017w4,karton2022quantum} is one of the few methods \cite{Bartlett:09,zahrt2019prediction,thorpe2024factorized}  at present  \cite{mcardle2020quantum} can reach this level of accuracy for general many-electron systems.

In the previous work, the PSA scheme has been extended to general orders \cite{wang2024general}. The fourth-order termination of the Baker–Campbell–Hausdorff expansion and only connected $HT$ contractions appearing in the equations are preserved.  In the present work, we aim at generating and implementing the working equations. To this end,  additional aspects such as linear combinations of  projection manifolds, a simple hash-table canonicalization method, and numerical results will be  discussed.

\section{Generating working equations}
\subsection{Linear combinations of projection manifolds}
The energy and amplitudes of a general-order PSA CC can be obtained from the solutions of the following equations \cite{wang2024general}
\begin{align}
\bra{\Phi}  H e^{{T}^{\mathrm{PSA}}  }  \ket{\Phi}_{\mathrm{c}} &= E  \label{cce_c} \\
\bra{\mu^{\mathrm{PSA}} } {H} e^{{T}^{\mathrm{PSA}} } \ket{\Phi}_{\mathrm{c} } &=   0  \label{ccr_c}
\end{align}
where $\bra{\mu^{\mathrm{PSA}} } := \{  \bra{\mu_1^{\mathrm{PSA}}}, \bra{\mu_2^{\mathrm{PSA}}}, \cdots \}$  and ${T}^{\mathrm{PSA}}$ are the projection manifolds and excitation operator defined previously \cite{wang2024general}.  $H$, $\Phi$, and  $_c$ stand for the many-electron Hamiltonian, reference high-spin Slater determinant, and connected terms \cite{wang2018simple}.

In addition, similar to the closed-shell methods\cite{pulay1984efficient,hampel1992comparison,koch1990coupled,wang2018simple}, it is expected that  linear combinations of the projection manifolds will reduce the number of equations and accelerate the convergence of CC calculations. Namely, the amplitude equations, Eq. \eqref{ccr_c}, become
\begin{align}
\bra{\check{{\mu}}^{\mathrm{PSA}} } {H} e^{{T}^{\mathrm{PSA}} } \ket{\Phi}_{\mathrm{c} } &=   0  \label{ccr_c_combine}
\end{align}
where  $\bra{\check{\mu}^{\mathrm{PSA}} }$ is a linear combination of the excitation operators on the reference bra vector $\bra{\Phi}$. $\bra{\check{\Phi}^{AB}_{IJ}} :=  \bra{\Phi} \check{E}_{AB}^{IJ} := \bra{\Phi} (\frac{1}{3} E_{AB}^{IJ} + \frac{1}{6} E_{AB}^{JI})$ is one example. Since the notation $\tilde{}$ has been used in the normal ordering \cite{Kutzelnigg:97}, the notation $\check{}$  was adopted. Under the PSA scheme, there is no difference between a normal ordered operator or not for the cluster operator $T^{\mathrm{PSA}}$ and the excitation manifold. Therefore, the excitation operator, e.g., $E_{AB}^{IJ}$, may not have a $\tilde{}$ notation on top of it.

Nevertheless, the construction the metric matrix $M$ \cite{pulay1984efficient,hampel1992comparison,koch1990coupled,wang2018simple} becomes increasingly laborious. At this stage, we found that using similar structure as in the closed-shell method would lead to a feasible convergence. For example:
\begin{align}
{\check{E}}_{A}^{I} &= \frac{1}{2} E_{A}^{I} \label{eab} \\
{\check{E}}_{A\alpha}^{V\alpha} &= E_{A\alpha}^{V\alpha} \\
{\check{E}}_{V\beta}^{I\beta} &= E_{V\beta}^{I\beta} \\
\check{E}_{AB}^{IJ} &= \frac{1}{3} E_{AB}^{IJ} + \frac{1}{6} E_{AB}^{JI} \\
\check{E}_{A\alpha B}^{V\alpha I} &=  4 ( \frac{1}{3} E_{A\alpha B}^{V\alpha I} + \frac{1}{6}  E_{A B\alpha}^{ I V\alpha} ) \label{eabvi} \\
\check{E}_{A\alpha B\alpha }^{V\alpha W\alpha } &= 3 ( \frac{1}{3} E_{A\alpha B\alpha}^{V\alpha W\alpha} + \frac{1}{6}  E_{A\alpha B\alpha }^{W\alpha V\alpha} ) \\
\check{E}_{AV\beta}^{IJ\beta} &= 4 ( \frac{1}{3} E_{A V\beta}^{I J\beta} + \frac{1}{6}  E_{A V\beta}^{ J I\beta} ) \\
\check{E}_{A\alpha W \beta}^{V\alpha I \beta} &= 6( \frac{1}{3} E_{A\alpha W \beta}^{V\alpha I \beta}  ) \label{semi} \\
\check{E}_{V\beta W\beta}^{I\beta J \beta} &= 3 ( \frac{1}{3} E_{V\beta W\beta}^{I\beta J\beta} + \frac{1}{6}  E_{V\beta W\beta}^{J\beta I\beta} )  \\
\check{E}^{IJK}_{ABC} &= \frac{17}{120}  E^{IJK}_{ABC}  - \frac{1}{120}  E^{IKJ}_{ABC} - \frac{1}{120}    E^{JIK}_{ABC}
\notag\\&- \frac{7}{120}    E^{JKI}_{ABC}      -\frac{7}{120}    E^{KIJ}_{ABC}   -\frac{1}{120}    E^{KJI}_{ABC} \label{t1} \\
\check{E}^{IJV\alpha}_{ABC\alpha} &= 6( \frac{17}{120}  E^{IJV\alpha}_{ABC\alpha}  - \frac{1}{120}  E^{IV\alpha J}_{AB\alpha C} - \frac{1}{120}    E^{JIV\alpha}_{ABC\alpha}
\notag\\&   -  \frac{7}{120}    E^{JV\alpha I}_{AB\alpha C}      -\frac{7}{120}    E^{V\alpha IJ}_{A\alpha BC}   -\frac{1}{120}    E^{V\alpha JI}_{A\alpha BC} ) \label{t2}
\end{align}

It may be worthwhile to notice that,
\begin{enumerate}
\item[(i)] we scale the prefactor to make terms associated with the perturbative corrections become $\pm 1 $. For instance, $\frac{1}{2}$ and $4$ in Eqs.~\eqref{eab} and~\eqref{eabvi} would yield
$f_A^a t_a^I - f^I_i t^i_A  + \cdots $ and the residual $\check{r}_{AB}^{VI} :=  \langle \check{\Phi}^{AB}_{VI}| H e^{T^{\mathrm{PSA}}} | \Phi \rangle_{\mathrm{c}} = - [f^{\alpha}]_v^V t_{AB}^{vI} - f^I_i t_{AB}^{Vi}
 +  f_B^a t_{Aa}^{VI}
 + f_A^a t_{aB}^{VI} + \cdots $ in Eq. ~\eqref{ccr_c}. Though upper case symbols are often used for spatial orbitals \cite{Kutzelnigg:97,Bartlett:09,wang2018simple,wang2024general}, lower case indices are adopted, such as dummy variables and generated working equations in the present work.
 For sectors  $\check{r}_{ABV}^{WXI}$ and $\check{r}_{AVW}^{XIJ}$, with the presence of spin adaptation at the $[1|2|3]$ level, $t_{ABV}^{WXI} \neq - t_{BAV}^{WXI}$  and $t_{AVW}^{XIJ} \neq - t_{AWV}^{XIJ} $. The prefactor is set to include, both $t_{AVW}^{XIJ}$ and $t_{AWV}^{XIJ}$ contributions, since the deviations of antisymmetry  can be regarded as corrections. A list up to PSA-CCSDTQ is provided in Table \ref{prefactor}.
\item[(ii)] for a semi-internal excitation, e.g., $E_{A\alpha W\beta }^{V\alpha I\beta}$, there is only a single term in the linear combination of the excitation manifolds\cite{Knowles:93}, $c_1 E_{A\alpha W\beta }^{V\alpha I\beta} + c_2 E_{A\alpha W\beta }^{I\alpha V\beta } = c_1 E_{A\alpha W\beta }^{V\alpha I\beta} =  \check{E}_{A\alpha W\beta }^{V\alpha I\beta}  $, since $E_{A\alpha W\beta }^{I\alpha V\beta} $ does not appear in the PSA \cite{Janssen:91, Knowles:93, knowles2000erratum,  grotendorst2000modern, neogrady1994spin, Neogrady:95,urban1997spin, wang2024general} or the general spin-adapted CC methods \cite{Janssen:91,herrmann2022correctly}.
\item[(iii)] in the R$[1|2]$ level of spin adaptation, the prefactor for the corrections of spin-adaptation  $-1$ in $\langle \Phi_{VI}^{AW}| = \langle \Phi | ( E_{A \alpha W \beta}^{V \alpha I \beta }  - E_{A\alpha}^{I\alpha} I_W^V )$ from Table II in Ref. \cite{wang2024general},
is determined without assuming any linear combination of excitation manifolds (derived in Appendix A in Ref. \cite{wang2024general}) or scaling factor for a single excitation, e.g., $1/2$ in Eq. (\ref{eab}).
When we use linear combinations of excitation manifolds including different scaling factors of excitation operator themselves, e.g.,
$ \check{E}_{A}^{I} = c_1 E_{A}^{I} = c_1 E_{A\alpha}^{I\alpha} + c_1 E_{A\beta}^{I\beta}$ and $\check{E}_{A \alpha W \beta}^{V \alpha I \beta } = c_2 E_{A \alpha W \beta}^{V \alpha I \beta }$ , the semi-internal excitation manifold becomes
$\langle \check{\Phi}^{AV}_{WI} | =  \langle \Phi |  \check{E}_{A \alpha W \beta}^{V \alpha I \beta } +  \check{c} \check{E}_{A\alpha}^{I\alpha} I_W^V$ with $\check{c} = - c_2/c_1$, where -1 comes from $\mu^{\mathrm{PSA}[1|2]}_2$ term from Table II in Ref. \cite{wang2024general}. Thus, $\check{c} = - 6 \times \frac{1}{3} / \frac{1}{2} = -4$. Here, $c_2 = 6 \times \frac{1}{3}$ and $c_1 = \frac{1}{2}$ come from Eqs. (\ref{semi})  and (\ref{eab}), respectively. 
\item[(iv)] in the R$[1|2|3]$ level of spin adaptation,  the term with the unit matrix $I$, for example,  $\langle \Phi_{VIJ}^{AWB}| = \langle \Phi | (  E_{A\alpha W\beta B}^{V \alpha I\beta J} - I_W^V E_{A\alpha B}^{I\alpha J} ) $  in Table II of Ref. \cite{wang2024general},  the linear combination of projection manifolds is generated according to two aspects in the present implementation:
\begin{itemize}
\item[(iva)] triple excitation: $\check{E}^{V\alpha I\beta  J}_{A\alpha W\beta B}  = c_1 E_{A\alpha W\beta B}^{V \alpha I\beta J} + c_2 E_{A\alpha W\beta B}^{V \alpha J\beta I} + c_3 E_{B\alpha W\beta A}^{V \alpha I\beta J} + c_4 E_{B\alpha W\beta A}^{V \alpha J\beta I}$;
 \item[(ivb)]    the term  $ E_{A\alpha B}^{I\alpha J}$ via the linear combination in  $ \check{E}_{A B}^{I J} = 1/3 E_{A\alpha B}^{I\alpha J} + 1/6 E_{A\alpha B}^{J\alpha I} + 1/3 E_{A\beta B}^{I\beta J} + 1/6 E_{A\beta B}^{J\beta I} $.
\end{itemize}
  The match of these two linear combinations (iva) and (ivb),  would affect the spin adaptation in $\langle \check{\Phi}_{VIJ}^{AWB} |$. To ensure the correctness of the spin adaptation, we use
\begin{align}
\check{E}^{V\alpha I\beta  J}_{A\alpha W\beta B} &= 4(  E^{V\alpha I\beta  J }_{A\alpha W\beta B}   + 1/2    E^{ V\alpha J\beta I}_{A\alpha W\beta B})  \label{123_1} 
\end{align}
,instead of  17/120, -1/120, ... coefficients in Eqs. (\ref{t1}) - (\ref{t2}). The factors $1/2$ in Eqs. (\ref{123_1}) reflects the relation between $1/3$ and $1/6$ in $\check{E}_{A B}^{I J}$. Similarly,
\begin{align}
\check{E}^{V\alpha I\beta J \beta}_{A\alpha W\beta X\beta } &= 3(  E^{V\alpha I\beta J \beta}_{A\alpha W\beta X\beta } + 1/2    E^{V\alpha J\beta I \beta}_{A\alpha W\beta X\beta } ) \label{123_2}   \\
\check{E}^{V\alpha W\alpha I \beta}_{A\alpha B\alpha X \beta } &= 3(  E^{V\alpha W\alpha I \beta}_{A\alpha B\alpha X \beta } + 1/2    E^{V\alpha W\alpha I \beta}_{B\alpha A\alpha X \beta } \label{123_3}  ) 
\end{align}
For simplicity, the same linear combinations of projection excitations are used for other levels of spin adaptations. 

In analogy with (iii), $\langle \check{\Phi}_{VIJ}^{AWB} | =  \langle \Phi | (  \check{E}_{A\alpha W\beta B}^{V \alpha I\beta J} + \check{c}   I_W^V \check{E}_{A\alpha B}^{I\alpha J} )  $, the coefficient $\check{c}$ here can be determined as $\check{c} = - 4 / \frac{1}{3} = - 12$.

In  $ \langle \check{\Phi}_{VIJ}^{AWX}| =  \langle \Phi | (  \check{E}_{A\alpha W\beta X\beta}^{V \alpha I\beta J\beta} + \check{c}_1 I_W^V \check{E}_{A\alpha X\beta}^{I\alpha J\beta} + \check{c}_2   I_X^V \check{E}_{A\alpha W\beta}^{J\alpha I\beta} )  $ and 
$\langle \check{\Phi}_{VWI}^{ABX} |= \langle \Phi | ( \check{E}_{A\alpha B \alpha X \beta}^{V \alpha W\alpha I\beta} + \check{c}_3 I_X^W \check{E}_{A\alpha B\alpha}^{V\alpha I\alpha}   + \check{c}_4  I_X^V \check{E}_{A\alpha B\alpha}^{I\alpha W\alpha}   ) $,  the coefficients can be found as $\check{c}_1 = \check{c}_2 = \check{c}_3 = \check{c}_4 = -3 / \frac{4}{3} = -\frac{9}{4}$.
\end{enumerate}

\begin{table}[htbp]
  \centering
  \caption{Overall prefactors in the linear combinations of the projection excitations. The convention is, for example, the factor $4$ under $\check{E}_{A\alpha B}^{V\alpha I}$ corresponds to the same factor in Eq. (\ref{eabvi}). }
    \begin{tabular}{llllllll}           \hline
${\check{E}}_{A}^{I} $  & ${\check{E}}_{A\alpha}^{V\alpha} $  &  ${\check{E}}_{V\beta}^{I\beta}$    \\ 
$\frac{1}{2}$   &   1 &   1    \\  
$\check{E}_{AB}^{IJ}$  & $\check{E}_{A\alpha B}^{V\alpha I}$  &  $\check{E}_{A\alpha B\alpha }^{V\alpha W\alpha } $  &  $\check{E}_{AV\beta}^{IJ\beta} $  &   $\check{E}_{A\alpha W \beta}^{V\alpha I \beta}$ &  $\check{E}_{V\beta W\beta}^{I\beta J \beta}$    \\
1 & 4 & 3 &  4 &  6 &  3       \\
$\check{E}^{IJK}_{ABC}$ &  $\check{E}^{IJK\beta}_{A B V\beta}$      &   $ \check{E}^{V\alpha I J}_{A\alpha B C}$   &  $\check{E}^{I J \beta K \beta}_{A V\beta W\beta}$  &  $\check{E}_{A\alpha W\beta B}^{V \alpha I\beta J}  $  &  $\check{E}^{V\alpha W\alpha I}_{A\alpha B\alpha C}$  &    $ \check{E}^{I\beta J \beta K \beta}_{V\beta W\beta X\beta} $  &  $\check{E}_{A\alpha W \beta X\beta}^{V \alpha I\beta J\beta} $ \\
1 & 6 & 6 & 15/2 &  4  &  15/2  &  10/3 &  3 \\
  $ \check{E}_{A\alpha B\alpha X \beta}^{V \alpha W\alpha I\beta}$&  $ \check{E}^{V\alpha W\alpha X\alpha}_{A\alpha B\alpha C\alpha}$    \\
 3 &  10/3   \\
$\check{E}^{IJKL}_{ABCD}$  &   $\check{E}^{V\alpha IJK}_{A\alpha BCD}$ &$ \check{E}^{V\alpha IJK}_{A\alpha BCD}$ &  $\check{E}^{I JK\beta L\beta}_{ABV\beta W\beta}$ &  $ \check{E}^{ V\alpha I\beta J K }_{A\alpha W\beta B C} $&  $ \check{E}^{V\alpha W\alpha IJ}_{A\alpha B\alpha CD}$ &  $\check{E}^{I J\beta K\beta L\beta}_{AV\beta W\beta X\beta}$ &   $ \check{E}_{A\alpha W\beta X\beta B}^{V\alpha I\beta J\beta K} $ \\
1 & 8 & 8  & 14 & 5040/107 &  14 & 28/3 & 42 \\
$ \check{E}_{V\alpha W\alpha I\beta J}^{A\alpha B\alpha X\beta C}$   &  $\check{E}_{A\alpha B\alpha C\alpha D}^{V\alpha W\alpha X\alpha I}$  &
$\check{E}_{V\beta W\beta X\beta Y\beta}^{I\beta J\beta K\beta L\beta } $ & $\check{E}_{A\alpha W\beta X\beta Y\beta}^{V\alpha I\beta J\beta K\beta} $  &  $\check{E}_{A\alpha B\alpha X\beta Y\beta}^{V\alpha W\alpha I\beta J\beta} $ &  $\check{E}_{A\alpha B\alpha C\alpha Y\beta}^{V\alpha W\alpha X\alpha I\beta} $  &  $\check{E}^{A\alpha B\alpha C\alpha D\alpha }_{V\alpha W\alpha X\alpha Y\alpha }$ \\
21/2& 28/3  & 35/12& 35/3&  35/2 &  35/3 &  35/12
   \\ \hline
\end{tabular}
  \label{prefactor}
\end{table}

\subsection{Hash-table canonicalization algorithm}

We notice that though the double-coset algorithm \cite{wang2018simple} can substantially reduce the computational cost over the half-naive canonicalization algorithm (generates symmetry equivalent equations and minimize the dummy indices). Nevertheless, the time-demanding feature of the half-naive algorithm is partially due to the large number of non-canonicalized equations, that are generated from the permutations in the direct evaluation of coupling coefficient (DECC) \cite{wang2018simple} algorithm. Those equations from DECC need to be canonicalized termwise. This generation scheme implies one could build a hash-table that
\begin{itemize}
    \item keys are the symmetry equivalent with minimized dummy indices equations without the prefactors,
    \item values are the canonicalized equations without the prefactors. The choice of the canonicalized equation can be any entry among the equivalent forms \cite{wang2018simple, lechner2023internally} and can be regarded as the equation with the minimal value \cite{wang2018simple} of the sorted indices.
\end{itemize}
Since the average computational complexity in searching over hash table is $O(1)$ \cite{cormen2022introduction}, the computational cost will be comparable with DECC.  Notice hash table techniques have been used in tensor storage \cite{manzer2017general}, eliminations of duplicated expressions in intermediates of factorizations (reducing from $O(N^2)$) \cite{lechner2023internally,lechner2024code}, and eliminations of equivalent expressions due to dummy indices \cite{lai2014framework}.  

This consideration leads to the present hash-table canonicalization algorithm. We can start from an empty set of the symmetry-equivalent equations and loop over all each uncanonicalized equations from a common tensor contraction, e.g., $\langle \Phi | E_{ABC}^{IJK} f_{r\alpha}^{s\alpha} E_{s\alpha}^{r\alpha}  t^{ijk}_{abc} E^{abc}_{ijk} | \Phi \rangle$. 

If an uncanonicalized equation does not belong to the existing symmetry-equivalent equations as the keys of the hash table, the uncanonicalized equation and its symmetry equivalent forms under minimized dummy indices are added to the hash table. Then, the uncanonicalized is returned to the canonical equation among the symmetry equivalent forms (the value of the hash table). If the uncanonicalized belongs to the keys of existing hash table, the equation will also be returned to the canonical form as the value of the hash table.

After the equations from DECC are converted into the canonical forms. The equations differ by prefactors can be merged. This can be done by forming another hash table that the keys are the equations without prefactors and the values are the prefactors. The merging steps are searching over the hash table and updated the prefactors.

The hash table in this algorithm corresponds to the dictionary structure  in the programming language Python. The bottleneck of this algorithm can be in generating the symmetry equivalent forms of the equations (besides total number of terms in the uncanonicalized equations). These are the individual symmetries of amplitude tensors, the permutations between amplitude tensors, and two-electron integrals symmetries \cite{wang2018simple}. Though the computational complexity is factorial with respect to the order of the CC expansion,  this complexity exists in generating the CC equations in the DECC scheme.

Thus, we expect the computational time of canonicalization in this hash-table approach is comparable with the generating of equations in DECC, if the canonicalization is performed over tensors $t^{\alpha}$ and $t^{\beta}$ according to the left-hand side of '+=' or '=' in Table I in Ref. \cite{wang2024general}. Here $t^{\alpha}$ and $t^{\beta}$ mean the spin component are $\alpha$ and $\beta$ in the amplitude tensors, respectively, e.g., $t_{A\alpha}^{I\alpha}$ belongs to $t^{\alpha}$. Expanding over right-hand-side of '+=' in Table I in Ref. \cite{wang2024general} can unnecessarily produce more individual tensors and additional permutation symmetries. 

In addition, if one further simplifies the generating of CC equations by topological equivalent terms \cite{evangelista2022automatic}, the canonicalizations are expected to be achieved inside each topological equivalent type.

\subsection{Generation of working equations}
\label{gen_eqns}
In this subsection, we outline the workflow in generating equations for the PSA-CC approach.

\begin{enumerate}
\item[(i)] Select a level of the CC method and spin adaptation
\item[(ii)] Generate terms from cluster, Hamiltonian, and projection operators
\begin{enumerate}
\item[(iia)] Generate terms in the cluster operator $T^{\mathrm{PSA}}$
\begin{itemize}
\item Generate individual excitation terms, $T^{\mathrm{PSA}}_k, k=1,2,\cdots$  according to the level of the CC method

e.g.,
$\{t_A^I  E_I^A , t^{V}_{A} E_{V\alpha}^{A\alpha} , t^{I}_{V} E_{I\beta}^{V\beta}  \cdots \}$
\item Modify the excitation operators according to the level of spin adaptation, e.g., in the T$[1|2]$ level of the spin adaptation, $t_A^I E_I^A $ is replaced to $\{ t^{I\alpha}_{A\alpha} {E_{I\alpha}^{A\alpha} ,t^{I\beta}_{A\beta} E_{I\beta}^{A\beta}}  \}$
\end{itemize}
\item[(iib)] Generate terms $\{ H \}$  in Hamiltonian , i.e.,
$\{ f_{Q \alpha}^{P \alpha} {E}^{Q\alpha}_{P\alpha} , f_{Q \beta}^{P \beta} {E}^{Q\beta}_{P\beta}, \frac{1}{2} W_{PQ}^{RS} {E}_{RS}^{PQ} \}$
\item[(iic)] Generate terms $\{ R \}$  in excitation manifolds
\begin{itemize}
\item Generate projection operators $\{ E \}$ according to the level of CC method, e.g., $\{ E_{A}^{I}, E^{V\alpha}_{A\alpha},   E^{I\beta}_{V\beta} \cdots \}$
\item Modify excitation operators according to the spin adaptation, e.g., in R$[1|2]$ spin adaptation, $E_A^I $ is replaced to $\{ {E^{I\alpha}_{A\alpha} ,E^{I\beta}_{A\beta}} \}$
\end{itemize}
\item[(iid)] Generate spin indices for the excitation operators, e.g., $E_A^I$ is denoted to $\{0,0\}$, $E_{A\alpha}^{ V\alpha}$ is denoted to $\{1,1\}$, and $E_{V\beta}^{ I\beta}$ is denoted to $\{-1,-1\}$
\item[(iie)] Generate  a list of tensor contraction operators as product of $\{R\} \{H\} \{T \}$, combine the equivalent terms, e.g., $E_I^A E_{I\beta}^{A\beta}$ and $E_{I\beta}^{A\beta} E_I^A $ (they commute in the PSA scheme), and update the prefactors after merging the equivalent terms. Here $\{T \}$ includes the product of the PSA excitation operators.  To reduce the memory consumption, we first form the product $\{R\} \{H\}$, then add excitation operator from $T^{\mathrm{PSA}}$ successfully. The vanishing contractions, e.g., $\langle \Phi| {E}^{Q\alpha}_{P\alpha} E_I^A E_J^B E_K^C | \Phi \rangle  $ will be screened out during adding the product of the excitation operators in $\{T \}$.
\end{enumerate}

\item[(iii)] Loop over the  $\{R\} \{H\} \{T \}$ terms, and compute the contractions in each $\{R\} \{H\} \{T \}$ 
\begin{itemize}
 \item[(iiia)] Evaluate the contractions according to DECC for open-shell systems. In each circle, the spin indices are collected and used for determining terms, $O_{\textrm{cs}}, O_{{\textrm{os-ss}}}$, and $O_{{\textrm{os-os}}}$ of contractions in  Eq. 47  in Ref. \cite{wang2024general}
 \item[(iiib)] Replace the Fock matrices, $\{f^{\alpha}, f^{\beta} \}$, to the form encoded the Brillouin  conditions, according to Eqs. (16) - (20) in Ref. \cite{wang2024general}$^a$. Here $f^{\alpha}$ and $f^{\beta}$ correspond to $f_{Q\alpha}^{P\alpha}$ and $f_{Q\beta}^{P\beta}$ in Ref. \cite{wang2024general} respectively.
 \item[(iiic)] Replace the excitation manifolds, $\{ r^{\alpha}, r^{\beta} \}$ to $\{ r \}$,  according to Tables II and III in Ref. \cite{wang2024general}. Here  $\{ r^{\alpha}, r^{\beta} \}$ means the excitation operator with spin $\alpha$ and $\beta$ components, e.g., the term including $E^{I\alpha}_{A\alpha}$ belongs to $r^{\alpha}$
  \item[(iiid)] Permute the results from contractions according to linear combinations of excitation manifolds$^b$
  \item[(iiie)] Canonicalize and merge the equivalent expressions after the canonicalization of the resulting equations after (iiid)
\end{itemize}
\item[(iv)] Merge equivalent expressions among all  $\{R\} \{H\} \{T \}$ terms. The canonicalization in (iiie) is for each $\{R\} \{H\} \{T \}$  contraction separately \cite{wang2018simple}. The terms  $\{ r^{\alpha} \}$ and $\{ r^{\beta} \}$ could generate equivalent $r[AI]$ expressions. 
\item[(v)] Add the expressions of the polarized amplitudes, $\{ t^{\alpha}, t^{\beta} \}$ to $\{ t \}$, according to Table I in Ref. \cite{wang2024general}
\item[(vi)] Collect the equations related to permutations of the residual indices and the prefactors associated with the permutations, in analogous with closed-shell CCSDT and CCSDTQ \cite{wang2018simple}. Canonicalizations have been applied to ensure all related entries are obtained$^c$.
\end{enumerate}
$^a$$\bar{f}_Q^P$ in Eqs. (16) and (17) in Ref. \cite{wang2024general} can be written in terms of two-electron integrals as in Ref. \cite{Knowles:93}. \\
$^b$To  ensure residuals and amplitudes have the same type of domains, that makes a CC solver straightforward, the permutations are carried out by first permuting the upper Roman indices of the excitation tensor of the indices in the excitation operators associated with the residuals, if this permutation does not change the type of domain. Namely for $\langle \Phi | E_{ABC\alpha}^{IJV\alpha}$, it may be denoted to a (eeecca)-type domains. Here e, c, and a are denoted to external ($ABCD$), closed ($IJKL$), and active ($VXWY$) domains \cite{wang2018simple, wang2024general}. A (1,0,2) permutation leads to $\langle \Phi | E_{ABC\alpha}^{JIV\alpha}$, that is still a (eeecca)-type domain.  Otherwise, the lower indices of the excitation tensor of the indices associated with the residuals will be permuted. For example, $\langle \Phi | E_{ABC\alpha}^{IJV\alpha}$ with a (0,2,1) permutation on the upper indices will lead to $\langle \Phi | E_{ABC\alpha}^{IV\alpha J}$, that is a (eeecac)-type domains, no longer  (eeecca). Permuting the lower indices leads to   $\langle \Phi | E_{ACB\alpha}^{IJV\alpha}$. This operation can be justified by the symmetry of the excitation operator \cite{Kutzelnigg:97,wang2024general}. If both permuting first half and second half of the indices will change the domain type, the prefactor of this permutation will be set to zero. \\
$^c$No permutations or canonicalizations in the expressions of $t^{\alpha}$ and $t^{\beta}$.

\section{Results and discussions}

 The restricted open-shell Hartree-Fock (ROHF) and full configuration interaction (FCI) \cite{knowles1984new} values  were obtained by the PySCF program package version 2.2.1 \cite{sun2015libcint, sun2018pyscf,sun2020recent}. The complete-activate space configuration interaction (CAS-CI) module \cite{sun2017general} was used to compute the FCI values with frozen-core approximations in the PySCF \cite{sun2017general}.
The spin-orbital calculations were performed by the NWChem program version 7.2.0\cite{hirata2003tensor,apra2020nwchem}. 
The PSA-CC equations and energies were obtained by local software. The PSA-CC equations are described in the supplementary material and will be provided in the ancillary files.

We first examine the convergence of the PSA schemes with respect to the FCI limit. For a lithium atom, the correlation energies (as the energy difference between the post-HF and the ROHF methods \cite{pople1975correlation,Helgaker:00book,Bartlett:09} with the present basis set) obtained from PSA-CCSDT and FCI are presented in Table \ref{tab6}. Atomic units (a.u.) are used.

Similar to the work by Berente, Szalay, and Gauss \cite{berente2002spin}, PSA-T$[1|2|3]$R$[1|2|3]$-CCSDT/cc-pCVDZ \cite{prascher2011gaussian} provides near the same energy as the FCI value. One additional digit better agreement with respect to the FCI value was obtained from  PSA-T$[1|2|11|3]$R$[1|2|3]$-CCSDT/cc-pCVDZ. The remaining deviation can be attributed to further spin adaptations, such as $T_1 T_2$ and $T_1^3$ levels in Table I and Appendix A in Ref. \cite{wang2024general}.

\begin{table}[htbp]
  \caption{Correlation energies (a.u.) of the $X ^2S$ state of a lithium atom  at levels of spin adaptations in CCSDT methods comparing with the FCI value. cc-pCVDZ basis set \cite{prascher2011gaussian} is used.}
    \begin{tabular}{lrrr} \hline
    Method             &       &        &   Correlation energy       \\ \hline
    PSA-T[1]R[1]-CCSDT &        &   &  -0.03360 38254 73  \\
    PSA-T$[1|2]$R$[1|2]$-CCSDT &      &  & -0.03362 09885 83 \\ 
    PSA-T$[1|2|11]$R$[1|2]$-CCSDT &        &  & -0.03362 09886 61 \\
    PSA-T$[1|2|3]$R$[1|2|3]$-CCSDT &        &        &     -0.03362 10131 75 \\
    PSA-T$[1|2|11|3]$R$[1|2|3]$-CCSDT &         &           &  -0.03362 10132 66  \\ 
    FCI &       &                                        &  -0.03362 10132 44 \\ \hline
    \end{tabular}
  \label{tab6}
\end{table}

We then compared the PSA-T$[1|2]$R$[1|2]$,  PSA-T$[1|2|11]$R$[1|2]$, and PSA-T$[1|2|3]$R$[1|2|3]$ levels of calculations with the spin-orbital approaches in Table \ref{tab_diatomics}.

\begin{table}[htbp]
  \centering
  \caption{Electronic energies (a.u.) at levels of CC expansions comparing with the FCI limits. Core electrons are frozen in all post-HF calculations. Bond lengths and term symbols are adopted from the National Institute of Standards and Technology (NIST) Chemistry WebBook \cite{nistchem}. }
    \begin{tabular}{lrrrrr}           \hline
  cc-pVDZ \cite{dunning1989gaussian,prascher2011gaussian} & BeH $X^2\Sigma^{+a}$   & BH $a ^3\Pi^b$    & CH $X^2 \Pi^c$   & NH $X^3 \Sigma^{-d}$   & OH $X^2 \Pi^e$\\ \hline
   ROHF                              & -15.1494 & -25.1106 & -38.2688 & -54.9596 & -75.3900 \\
   PSA-T$[1|2]$R$[1|2]$-CCSD         & -0.03837 & -0.05573 & -0.10901 & -0.12998 & -0.16741 \\
   PSA-T$[1|2|11]$R$[1|2]$-CCSD      & -0.03837 & -0.05573 & -0.10901 & -0.12998 & -0.16741 \\
   PSA-T$[1|2]$R$[1|2]$-CCSDT        & -0.03900 & -0.05690 & -0.11138 & -0.13178 & -0.16929 \\
   PSA-T$[1|2|3]$R$[1|2|3]$-CCSDT    & -0.03901 & -0.05695 & -0.11142 & -0.13189 & -0.16934 \\
   PSA-T$[1|2]$R$[1|2]$-CCSDTQ       & -0.03900 & -0.05692 & -0.11145 & -0.13192 & -0.16956 \\
   PSA-T$[1|2|3]$R$[1|2|3]$-CCSDTQ   & -0.03901 & -0.05697 & -0.11150 & -0.13202 & -0.16961  \\
   Spin-orbital CCSD                  & -0.03838 & -0.05578 & -0.10907 & -0.13010 & -0.16747 \\
   Spin-orbital CCSDT                 & -0.03901 & -0.05696 & -0.11145 & -0.13197 & -0.16942 \\
   Spin-orbital CCSDTQ                & -0.03901 & -0.05697 & -0.11152 & -0.13209 & -0.16968 \\
   FCI                               & -0.03901 & -0.05697 & -0.11153 & -0.13210 & -0.16968 \\ \hline
\end{tabular}
\\
\hspace{-10cm} $^ar_{\textrm{BeH}}$ = 1.3426 \AA \\
\hspace{-10cm} $^br_{\textrm{BH}}$ = 1.2006 \AA  \\
\hspace{-10cm} $^cr_{\textrm{CH}}$ = 1.1199 \AA  \\
\hspace{-10cm} $^dr_{\textrm{NH}}$ = 1.0362 \AA  \\
\hspace{-10cm} $^er_{\textrm{OH}}$ = 0.9697 \AA \\
  \label{tab_diatomics}
\end{table}

As suggested by the numerical results in Table \ref{tab_diatomics}, the spin-orbital-based approaches converge to FCI faster than the present PSA levels. The energy differences between PSA and spin-orbital methods are about $10^{-4}$ to $10^{-5}$ a.u., which are below 0.1 kcal mol$^{-1}$.  The effects of T$[11]$ spin adaptation are negligible in the present scope.

We further notice the differences between PSA and spin-orbital CC are mainly at the CCSD levels. For example, in the OH molecule, the difference between the PSA-T$[1|2]$R$[1|2]$-CCSD and the spin-orbital CCSD is about 6 $\times$ 10$^{-5}$ a.u., that is similar to the difference between the PSA-T$[1|2|3]$R$[1|2|3]$-CCSDT and the spin-orbital CCSDT, 8 $\times$ 10$^{-5}$ a.u.. Likewise, the difference between PSA-T$[1|2|3]$R$[1|2|3]$-CCSDTQ and spin-orbital CCSDTQ is about 7 $\times$ 10$^{-5}$ a.u. These data suggest further improving the PSA accuracy may be more effective to increase the spin adaptations  at the CCSD levels, e.g., the T$[2|2]$ level, than the higher-order CC expansions such as CCSDT and CCSDTQ. This is similar to the compound methods \cite{martin1999towards,tajti2004heat,
yang2014ab,karton2017w4,karton2022quantum}. Namely, smaller basis sets with higher-order CC levels are effective to obtain results beyond the chemical accuracy. Since the size consistency issue \cite{hirata2015mutual} is practically unimportant at the PSA-T$[1|2]$R$[1|2]$-CCSD level \cite{Heckert:06}, this aspect may not be effective in the higher-order methods.

In addition, a number of potential improvement may be anticipated. The usage of permutations on the residual indices in step (vi) in Subsection \ref{gen_eqns} is a posterior collection. It may be feasible to be performed as a prior with the step (iiid) of using linear combinations of excitation operators. That would lead to a simplification of the workflow. Adopting orthogonal relations to accelerate convergence for different spin states \cite{knowles2000erratum} may be beneficial for general-order PSA-CC methods.

The number of resulting equations can be large at higher order PSA-CC and spin adaptations. This could slow down the computations. Though for a given basis set, the finite dimensional vector space for the computations from the spatial orbitals is smaller than from the spin orbitals. One may then expect a further optimized PSA-CC will be more efficient than the spin-orbital approaches. Nevertheless, it may be necessary to advance further improvement including factorizations \cite{kucharski1991recursive,kallay2001higher,hirata2003tensor,parkhill2010sparse,lai2012effective,pfeifer2014faster,manzer2017general,engels2011fully}. For example, in the present PSA-CC, there are many contractions involving active indices
\begin{align}
 r[\,] &+=  W[aijv] \,\, t[avji] \label{contract_1} \\
 r[\,] &+= - \frac{1}{2} W[aijv] \,\, t[avij]    \label{contract_2}
\end{align}
Since the number of active orbitals (1 occupation in the reference determinant) is typically smaller than the inactive orbitals (0 or 2 occupation in the reference determinant), it may be more efficient to combine multiple tensors into larger arrays:
\begin{align}
\mathbf{v}_1 &:= ( W[aijv], -\frac{1}{2} W[aijv]) \\
\mathbf{v}_2 &:= \begin{pmatrix}
           t[avji] \\
           t[avij]  
         \end{pmatrix}
\end{align}
and express the result of Eqs. (\ref{contract_1}) and (\ref{contract_2}) as $\mathbf{v}_1 \cdot \mathbf{v}_2$. Thus, the arrays with the small active indices in dimensions may be merged as a larger dimension. 
Protocol tensor contraction examples with a comparison of timing using NumPy\cite{harris2020array} and opt\_einsum \cite{daniel2018opt} packages are described in the supplementary materials and the Python file will be provided in the ancillary file. The results suggest that the merging approach can be more efficient.

\section{Summary}

In the present work, we reported  the further formulations for the linear combinations of the projection manifolds, the hash-table canonicalization algorithm, and the numerical results for the previous general-order open-shell CC method based on the PSA scheme \cite{wang2024general}. The energy differences between the present approaches and the spin-orbital CC methods are below 0.1 kcal mol$^{-1}$. After further optimizations the PSA-CC methods such as factorizations and merging equations with active indices, the current approach is expected to be an improvement with lower prefactors of computational costs for the spin-orbital based general-order CC methods for open-shell systems.

\section*{Acknowledgment}

This work was supported by a starting grant of Pennsylvania State University when the author was in the group of Professor Knizia.
We thank Professor Knizia for the comments, discussions, and supports. We thank Professors K\'{a}llay and Knowles for helpful discussions.

\section*{References}

\bibliographystyle{pccp}
\bibliography{topic_os}

\begin{thebibliography}{100}

\bibitem{bartlett1995coupled}
R.~J. Bartlett in {\em Modern Electronic Structure Theory, Part II}, ed. D.~R. Yarkony;
\newblock World Scientific, Singapore, 1995;
\newblock pp. 1047--1131.

\bibitem{christiansen2006coupled}
O.~Christiansen, {\em Theor. Chem. Acc.}, 2006, {\bf 116}, 106--123.

\bibitem{Bartlett:07}
R.~J. Bartlett and M.~Musia{\l}, {\em Rev. Mod. Phys.}, 2007, {\bf 79}, 291--352.

\bibitem{crawford2007introduction}
T.~D. Crawford and H.~F. Schaefer~III, {\em Rev. Comp. Chem.}, 2007, {\bf 14}, 33--136.

\bibitem{Bartlett:09}
I.~Shavitt and R.~J. Bartlett, {\em Many-Body Methods in Chemistry and Physics: MBPT and Coupled-Cluster Theory}, Cambridge University Press, Cambridge, 2009.

\bibitem{bartlett2012coupled}
R.~J. Bartlett, {\em Wiley Interdiscip. Rev. Comput. Mol. Sci.}, 2012, {\bf 2}, 126--138.

\bibitem{Lyakh:12}
D.~I. Lyakh, M.~Musiał, V.~F. Lotrich, and R.~J. Bartlett, {\em Chem. Rev.}, 2012, {\bf 112}, 182--243.

\bibitem{bartlett2024perspective}
R.~J. Bartlett, {\em Phys. Chem. Chem. Phys.}, 2024, {\bf 26}, 8013--8037.

\bibitem{Helgaker:00book}
T.~Helgaker, P.~J{\o}rgensen, and J.~Olsen, {\em Molecular Electronic-Structure Theory}, Wiley, Chichester, 2000.

\bibitem{Harris:99}
F.~E. Harris, {\em Int. J. Quantum Chem.}, 1999, {\bf 75}, 593--597.

\bibitem{Olsen:00}
J.~Olsen, {\em J. Chem. Phys.}, 2000, {\bf 113}, 7140--7148.

\bibitem{Hirata:00}
S.~Hirata and R.~J. Bartlett, {\em Chem. Phys. Lett.}, 2000, {\bf 321}, 216–224.

\bibitem{hirata2003tensor}
S.~Hirata, {\em J. Phys. Chem. A}, 2003, {\bf 107}, 9887--9897.

\bibitem{Kallay:00}
M.~Kall{\'a}y and P.~R. Surj{\'a}n, {\em J. Chem. Phys.}, 2000, {\bf 113}, 1359--1365.

\bibitem{kallay2001higher}
M.~K{\'a}llay and P.~R. Surj{\'a}n, {\em J. Chem. Phys.}, 2001, {\bf 115}, 2945--2954.

\bibitem{auer2006automatic}
A.~A. Auer, G.~Baumgartner, D.~E. Bernholdt, A.~Bibireata, V.~Choppella, D.~Cociorva, X.~Gao, R.~Harrison, S.~Krishnamoorthy, S.~Krishnan, C.-C. Lam, Q.~Lu, M.~Nooijen, R.~Pitzer, J.~Ramanujam, P.~Sadayappan, and A.~Sibiryakov, {\em Mol. Phys.}, 2006, {\bf 104}, 211--228.

\bibitem{parkhill2010sparse}
J.~A. Parkhill and M.~Head-Gordon, {\em Mol. Phys.}, 2010, {\bf 108}, 513--522.

\bibitem{engels2011fully}
A.~Engels-Putzka and M.~Hanrath, {\em J. Chem. Phys.}, 2011, {\bf 134}, 124106.

\bibitem{evangelista2022automatic}
F.~A. Evangelista, {\em J. Chem. Phys.}, 2022, {\bf 157}, 064111.

\bibitem{Stanton:14}
D.~A. Matthew, J.~Gauss, and J.~F. Stanton, {\em J. Chem. Theory Comput.}, 2013, {\bf 9}, 2567--2572.

\bibitem{Stanton:15}
D.~A. Matthew and J.~F. Stanton, {\em J. Chem. Phys.}, 2015, {\bf 142}, 064108.

\bibitem{scuseria1987closed}
G.~E. Scuseria, A.~C. Scheiner, T.~J. Lee, J.~E. Rice, and H.~F. Schaefer~III, {\em J. Chem. Phys.}, 1987, {\bf 86}, 2881--2890.

\bibitem{scuseria1988efficient}
G.~E. Scuseria, C.~L. Janssen, and H.~F. Schaefer, {\em J. Chem. Phys.}, 1988, {\bf 89}(12), 7382--7387.

\bibitem{lee1988efficient}
T.~J. Lee and J.~E. Rice, {\em Chem. Phys. Lett.}, 1988, {\bf 150}, 406--415.

\bibitem{koch1990coupled}
H.~Koch, H.~J.~A. Jensen, P.~J{\o}rgensen, T.~Helgaker, G.~E. Scuseria, and H.~F. Schaefer~III, {\em J. Chem. Phys.}, 1990, {\bf 92}, 4924--4940.

\bibitem{hampel1992comparison}
C.~Hampel, K.~A. Peterson, and H.-J. Werner, {\em Chem. Phys. Lett.}, 1992, {\bf 190}, 1--12.

\bibitem{matthews2014non}
D.~A. Matthews {\em Non-orthogonal spin-adaptation and application to coupled cluster up to quadruple excitations} PhD thesis, The University of Texas at Austin, 2014.

\bibitem{wang2018simple}
C.~Wang and G.~Knizia, {\em arXiv preprint arXiv:1805.00565}, 2018.

\bibitem{matthews2019diagrams}
D.~A. Matthews and J.~F. Stanton in {\em Mathematical Physics in Theoretical Chemistry}, ed. S.~Blinder and J.~House;
\newblock Elsevier, Amsterdam, 2019;
\newblock pp. 327--375.

\bibitem{lindgren1978coupled}
I.~Lindgren, {\em Int. J. Quantum Chem.}, 1978, {\bf 14}, 33--58.

\bibitem{nakatsuji1977cluster}
H.~Nakatsuji and K.~Hirao, {\em Chem. Phys. Lett.}, 1977, {\bf 47}, 569--571.

\bibitem{nakatsuji1978cluster_a}
H.~Nakatsuji and K.~Hirao, {\em J. Chem. Phys.}, 1978, {\bf 68}, 2053--2065.

\bibitem{nakatsuji1978cluster}
H.~Nakatsuji, {\em Chem. Phys. Lett.}, 1978, {\bf 59}, 362--364.

\bibitem{mukherjee1979hierarchy}
D.~Mukherjee, {\em Pramana}, 1979, {\bf 12}, 203--225.

\bibitem{nakatsuji1979cluster}
H.~Nakatsuji, {\em Chem. Phys. Lett.}, 1979, {\bf 67}, 329--333.

\bibitem{nakatsuji1983cluster}
H.~Nakatsuji, K.~Ohta, and T.~Yonezawa, {\em J. Phys. Chem.}, 1983, {\bf 87}, 3068--3074.

\bibitem{haque1984application}
M.~A. Haque and D.~Mukherjee, {\em J. Chem. Phys.}, 1984, {\bf 80}, 5058--5069.

\bibitem{mukherjee1989use}
D.~Mukherjee and S.~Pal, {\em Adv. Quantum Chem.}, 1989, {\bf 20}, 291--373.

\bibitem{Janssen:91}
C.~Janssen and H.~F. {Schaefer III}, {\em Theor. Chim. Acta}, 1991, {\bf 79}, 1--42.

\bibitem{Knowles:93}
P.~J. Knowles, C.~Hampel, and H.-J. Werner, {\em J. Chem. Phys.}, 1993, {\bf 99}, 5219--5227.

\bibitem{jayatilaka1993open}
D.~Jayatilaka and T.~J. Lee, {\em J. Chem. Phys.}, 1993, {\bf 98}, 9734--9747.

\bibitem{Li:94}
X.~Li and J.~Paldus, {\em J. Chem. Phys.}, 1994, {\bf 101}, 8812--8826.

\bibitem{neogrady1994spin}
P.~Neogr{\'a}dy, M.~Urban, and I.~Huba\v{c}, {\em J. Chem. Phys.}, 1994, {\bf 100}, 3706--3716.

\bibitem{Jeziorski:95a}
B.~Jeziorski, J.~Paldus, and P.~Jankowski, {\em Int. J. Quantum Chem.}, 1995, {\bf 56}, 129--155.

\bibitem{Neogrady:95}
P.~Neogr\'{a}dy and M.~Urban, {\em Int. J. Quantum Chem.}, 1995, {\bf 55}, 187--203.

\bibitem{nooijen1996many}
M.~Nooijen, {\em J. Chem. Phys.}, 1996, {\bf 104}, 2638--2651.

\bibitem{nooijen1996general}
M.~Nooijen and R.~J. Bartlett, {\em J. Chem. Phys.}, 1996, {\bf 104}, 2652--2668.

\bibitem{urban1997spin}
M.~Urban, P.~Neogr{\'a}dy, and I.~Huba{\v{c}} in {\em Recent Advances In Coupled-Cluster Methods}, ed. R.~J. Bartlett;
\newblock World Scientific, Singapore, 1997;
\newblock pp. 275--306.

\bibitem{Szalay:97}
P.~G. Szalay and J.~Gauss, {\em J. Chem. Phys.}, 1997, {\bf 107}, 9028--9038.

\bibitem{Szalay:00}
P.~G. Szalay and J.~Gauss, {\em J. Chem. Phys.}, 2000, {\bf 112}, 4027--4036.

\bibitem{Li:98}
X.~Li and J.~Paldus, {\em Mol. Phys.}, 1998, {\bf 94}, 41--54.

\bibitem{Jeziorski:99}
P.~Jankowski and B.~Jeziorski, {\em J. Chem. Phys.}, 1999, {\bf 111}, 1857--1869.

\bibitem{nooijen2001towards}
M.~Nooijen and V.~Lotrich, {\em J. Mol. Struct.: THEOCHEM}, 2001, {\bf 547}, 253--267.

\bibitem{grotendorst2000modern}
P.~Knowles, M.~Sch{\"u}tz, and H.-J. Werner in {\em Modern Methods and Algorithms of Quantum Chemistry}, ed. J.~Grotendorst, Vol. ~1;
\newblock John von Neumann Institute for Computing, J\"{u}lich, 2000;
\newblock pp. 69--151.

\bibitem{knowles2000erratum}
P.~J. Knowles, C.~Hampel, and H.-J. Werner, {\em J. Chem. Phys.}, 2000, {\bf 112}, 3106--3107.

\bibitem{Heckert:06}
M.~Heckert, O.~Heun, J.~Gauss, and P.~G. Szalay, {\em J. Chem. Phys.}, 2006, {\bf 124}, 124105.

\bibitem{Datta:08}
D.~Datta and D.~Mukherjee, {\em Int. J. Quantum Chem.}, 2008, {\bf 108}, 2211--2222.

\bibitem{datta2011state}
D.~Datta, L.~Kong, and M.~Nooijen, {\em J. Chem. Phys.}, 2011, {\bf 134}, 214116.

\bibitem{Datta:13}
D.~Datta and J.~Gauss, {\em J. Chem. Theory Comput.}, 2013, {\bf 9}, 2639--2653.

\bibitem{datta2014analytic}
D.~Datta and J.~Gauss, {\em J. Chem. Phys.}, 2014, {\bf 141}, 104102.

\bibitem{gunasekera2024multireference}
A.~D. Gunasekera, N.~Lee, and D.~P. Tew, {\em Faraday Discuss.}, 2024, {\bf 254}, 170--190.

\bibitem{datta2015communication}
D.~Datta and J.~Gauss, {\em J. Chem. Phys.}, 2015, {\bf 143}, 011101.

\bibitem{datta2019accurate}
D.~Datta and J.~Gauss, {\em J. Chem. Theory Comput.}, 2019, {\bf 15}, 1572--1592.

\bibitem{herrmann2020generation}
N.~Herrmann and M.~Hanrath, {\em J. Chem. Phys.}, 2020, {\bf 153}, 164114.

\bibitem{herrmann2022analysis}
N.~Herrmann and M.~Hanrath, {\em Mol. Phys.}, 2022, {\bf 120}, e2005836.

\bibitem{herrmann2022correctly}
N.~Herrmann and M.~Hanrath, {\em J. Chem. Phys.}, 2022, {\bf 156}, 054111.

\bibitem{folkestad2023entanglement}
S.~D. Folkestad, B.~S. Sannes, and H.~Koch, {\em J. Chem. Phys.}, 2023, {\bf 158}, 224114.

\bibitem{krieger2021impact}
A.~M. Krieger and E.~A. Pidko, {\em ChemCatChem}, 2021, {\bf 13}, 3517--3524.

\bibitem{laplaza2024overcoming}
R.~Laplaza, M.~D. Wodrich, and C.~Corminboeuf, {\em J. Phys. Chem. Lett}, 2024, {\bf 15}, 7363--7370.

\bibitem{martin1999towards}
J.~M. Martin and G.~de~Oliveira, {\em J. Chem. Phys.}, 1999, {\bf 111}, 1843--1856.

\bibitem{tajti2004heat}
A.~Tajti, P.~G. Szalay, A.~G. Cs{\'a}sz{\'a}r, M.~K{\'a}llay, J.~Gauss, E.~F. Valeev, B.~A. Flowers, J.~V{\'a}zquez, and J.~F. Stanton, {\em J. Chem. Phys.}, 2004, {\bf 121}, 11599--11613.

\bibitem{yang2014ab}
J.~Yang, W.~Hu, D.~Usvyat, D.~Matthews, M.~Sch{\"u}tz, and G.~K.-L. Chan, {\em Science}, 2014, {\bf 345}, 640--643.

\bibitem{karton2017w4}
A.~Karton, N.~Sylvetsky, and J.~M. Martin, {\em J. Comput. Chem.}, 2017, {\bf 38}, 2063--2075.

\bibitem{karton2022quantum}
A.~Karton in {\em Annu. Rep. Comput. Chem}, ed. D.~A. Dixon, Vol. ~18;
\newblock Elsevier, Amsterdam, 2022;
\newblock pp. 123--166.

\bibitem{zahrt2019prediction}
A.~F. Zahrt, J.~J. Henle, B.~T. Rose, Y.~Wang, W.~T. Darrow, and S.~E. Denmark, {\em Science}, 2019, {\bf 363}, eaau5631.

\bibitem{thorpe2024factorized}
J.~H. Thorpe, Z.~W. Windom, R.~J. Bartlett, and D.~A. Matthews, {\em J. Phys. Chem. A}, 2024, {\bf 128}, 7720--7732.

\bibitem{mcardle2020quantum}
S.~McArdle, S.~Endo, A.~Aspuru-Guzik, S.~C. Benjamin, and X.~Yuan, {\em Rev. Mod. Phys.}, 2020, {\bf 92}, 015003.

\bibitem{wang2024general}
C.~Wang, {\em arXiv preprint arXiv:2403.10128}, 2024.

\bibitem{pulay1984efficient}
P.~Pulay, S.~Saeb{\o}, and W.~Meyer, {\em J. Chem. Phys.}, 1984, {\bf 81}, 1901--1905.

\bibitem{Kutzelnigg:97}
W.~Kutzelnigg and D.~Mukherjee, {\em J. Chem. Phys.}, 1997, {\bf 107}, 432--449.

\bibitem{lechner2023internally}
M.~H. Lechner {\em Internally Contracted Multireference Coupled-Cluster Theories With Automated Code Generation} PhD thesis, Universit{\"a}ts-und Landesbibliothek Bonn, 2023.

\bibitem{cormen2022introduction}
T.~H. Cormen, C.~E. Leiserson, R.~L. Rivest, and C.~Stein, {\em Introduction to algorithms}, MIT press, 2022.

\bibitem{manzer2017general}
S.~Manzer, E.~Epifanovsky, A.~I. Krylov, and M.~Head-Gordon, {\em J. Chem. Theory Comput.}, 2017, {\bf 13}, 1108--1116.

\bibitem{lechner2024code}
M.~H. Lechner, A.~Papadopoulos, K.~Sivalingam, A.~A. Auer, A.~Koslowski, U.~Becker, F.~Wennmohs, and F.~Neese, {\em Phys. Chem. Chem. Phys.}, 2024, {\bf 26}, 15205--15220.

\bibitem{lai2014framework}
P.-W. Lai, {\em A Framework for Performance Optimization of Tensor Contraction Expressions}, The Ohio State University, 2014.

\bibitem{knowles1984new}
P.~J. Knowles and N.~C. Handy, {\em Chem. Phys. Lett.}, 1984, {\bf 111}, 315--321.

\bibitem{sun2015libcint}
Q.~Sun, {\em J. Comp. Chem.}, 2015, {\bf 36}, 1664--1671.

\bibitem{sun2018pyscf}
Q.~Sun, T.~C. Berkelbach, N.~S. Blunt, G.~H. Booth, S.~Guo, Z.~Li, J.~Liu, J.~D. McClain, E.~R. Sayfutyarova, S.~Sharma, et~al., {\em WIREs Comput. Mol. Sci.}, 2018, {\bf 8}, e1340.

\bibitem{sun2020recent}
Q.~Sun, X.~Zhang, S.~Banerjee, P.~Bao, M.~Barbry, N.~S. Blunt, N.~A. Bogdanov, G.~H. Booth, J.~Chen, Z.-H. Cui, J.~J. Eriksen, Y.~Gao, S.~Guo, J.~Hermann, M.~R. Hermes, K.~Koh, P.~Koval, S.~Lehtola, Z.~Li, J.~Liu, N.~Mardirossian, J.~D. McClain, M.~Motta, B.~Mussard, H.~Q. Pham, A.~Pulkin, W.~Purwanto, P.~J. Robinson, E.~Ronca, E.~R. Sayfutyarova, M.~Scheurer, H.~F. Schurkus, J.~E.~T. Smith, C.~Sun, S.-N. Sun, S.~Upadhyay, L.~K. Wagner, X.~Wang, A.~White, J.~D. Whitfield, M.~J. Williamson, S.~Wouters, J.~Yang, J.~M. Yu, T.~Zhu, T.~C. Berkelbach, S.~Sharma, A.~Y. Sokolov, and G.~K.-L. Chan, {\em J. Chem. Phys.}, 2020, {\bf 153}, 024109.

\bibitem{sun2017general}
Q.~Sun, J.~Yang, and G.~K.-L. Chan, {\em Chem. Phys. Lett.}, 2017, {\bf 683}, 291--299.

\bibitem{apra2020nwchem}
E.~Apr\`{a}, E.~J. Bylaska, W.~A. de~Jong, N.~Govind, K.~Kowalski, T.~P. Straatsma, M.~Valiev, H.~J.~J. van Dam, Y.~Alexeev, J.~Anchell, V.~Anisimov, F.~W. Aquino, R.~Atta-Fynn, J.~Autschbach, N.~P. Bauman, J.~C. Becca, D.~E. Bernholdt, K.~Bhaskaran-Nair, S.~Bogatko, P.~Borowski, J.~Boschen, J.~Brabec, A.~Bruner, E.~Cau\"{e}t, Y.~Chen, G.~N. Chuev, C.~J. Cramer, J.~Daily, M.~J.~O. Deegan, J.~Dunning, T.~H., M.~Dupuis, K.~G. Dyall, G.~I. Fann, S.~A. Fischer, A.~Fonari, H.~Fr\"{u}chtl, L.~Gagliardi, J.~Garza, N.~Gawande, S.~Ghosh, K.~Glaesemann, A.~W. G\"{o}tz, J.~Hammond, V.~Helms, E.~D. Hermes, K.~Hirao, S.~Hirata, M.~Jacquelin, L.~Jensen, B.~G. Johnson, H.~J\'{o}nsson, R.~A. Kendall, M.~Klemm, R.~Kobayashi, V.~Konkov, S.~Krishnamoorthy, M.~Krishnan, Z.~Lin, R.~D. Lins, R.~J. Littlefield, A.~J. Logsdail, K.~Lopata, W.~Ma, A.~V. Marenich, J.~Martin~del Campo, D.~Mejia-Rodriguez, J.~E. Moore, J.~M. Mullin, T.~Nakajima, D.~R. Nascimento, J.~A. Nichols, P.~J. Nichols, J.~Nieplocha, A.~Otero-de-la Roza, B.~Palmer,
  A.~Panyala, T.~Pirojsirikul, B.~Peng, R.~Peverati, J.~Pittner, L.~Pollack, R.~M. Richard, P.~Sadayappan, G.~C. Schatz, W.~A. Shelton, D.~W. Silverstein, D.~M.~A. Smith, T.~A. Soares, D.~Song, M.~Swart, H.~L. Taylor, G.~S. Thomas, V.~Tipparaju, D.~G. Truhlar, K.~Tsemekhman, T.~Van~Voorhis, A.~Vázquez-Mayagoitia, P.~Verma, O.~Villa, A.~Vishnu, K.~D. Vogiatzis, D.~Wang, J.~H. Weare, M.~J. Williamson, T.~L. Windus, K.~Woliński, A.~T. Wong, Q.~Wu, C.~Yang, Q.~Yu, M.~Zacharias, Z.~Zhang, Y.~Zhao, and R.~J. Harrison, {\em J. Chem. Phys.}, 2020, {\bf 152}, 184102.

\bibitem{pople1975correlation}
J.~Pople and J.~Binkley, {\em Mol. Phys.}, 1975, {\bf 29}, 599--611.

\bibitem{berente2002spin}
I.~Berente, P.~G. Szalay, and J.~Gauss, {\em J. Chem. Phys.}, 2002, {\bf 117}, 7872--7881.

\bibitem{prascher2011gaussian}
B.~P. Prascher, D.~E. Woon, K.~A. Peterson, T.~H. Dunning, and A.~K. Wilson, {\em Theore. Chem. Acc.}, 2011, {\bf 128}, 69--82.

\bibitem{nistchem}
K.~P. Huber and G.~H. Herzberg in {\em NIST Chemistry WebBook, NIST Standard Reference Database Number 69}, ed. P.~Linstrom and W.~Mallard;
\newblock National Institute of Standards and Technology, Gaithersburg MD, 20899, retrieved November 30, 2024.

\bibitem{dunning1989gaussian}
T.~H. Dunning~Jr, {\em J. Chem. Phys.}, 1989, {\bf 90}, 1007--1023.

\bibitem{hirata2015mutual}
S.~Hirata and I.~Grabowski in {\em Thom H. Dunning, Jr. A Festschrift from Theoretical Chemistry Accounts}, ed. A.~K. Wilson, K.~A. Peterson, and D.~E. Woon;
\newblock Springer, 2015;
\newblock pp. 267--275.

\bibitem{kucharski1991recursive}
S.~A. Kucharski and R.~J. Bartlett, {\em Theor. Chim. Acta}, 1991, {\bf 80}, 387--405.

\bibitem{lai2012effective}
P.-W. Lai, H.~Zhang, S.~Rajbhandari, E.~Valeev, K.~Kowalski, and P.~Sadayappan, {\em Procedia Comput. Sci.}, 2012, {\bf 9}, 412--421.

\bibitem{pfeifer2014faster}
R.~N. Pfeifer, J.~Haegeman, and F.~Verstraete, {\em Phys. Rev. E}, 2014, {\bf 90}, 033315.

\bibitem{harris2020array}
C.~R. Harris, K.~J. Millman, S.~J. van~der Walt, R.~Gommers, P.~Virtanen, D.~Cournapeau, E.~Wieser, J.~Taylor, S.~Berg, N.~J. Smith, R.~Kern, M.~Picus, S.~Hoyer, M.~H. van Kerkwijk, M.~Brett, A.~Haldane, J.~F. del R{\'{i}}o, M.~Wiebe, P.~Peterson, P.~G{\'{e}}rard-Marchant, K.~Sheppard, T.~Reddy, W.~Weckesser, H.~Abbasi, C.~Gohlke, and T.~E. Oliphant, {\em Nature}, 2020, {\bf 585}, 357--362.

\bibitem{daniel2018opt}
G.~Daniel and J.~Gray, {\em J. Open Source Softw.}, 2018, {\bf 3}, 753.

\end{thebibliography}


\begin{thebibliography}{1}

\bibitem{wang2018simple}
C.~Wang and G.~Knizia, {\em arXiv preprint arXiv:1805.00565}, 2018.

\bibitem{daniel2018opt}
G.~Daniel and J.~Gray, {\em J. Open Source Softw.}, 2018, {\bf 3}, 753.

\bibitem{harris2020array}
C.~R. Harris, K.~J. Millman, S.~J. van~der Walt, R.~Gommers, P.~Virtanen, D.~Cournapeau, E.~Wieser, J.~Taylor, S.~Berg, N.~J. Smith, R.~Kern, M.~Picus, S.~Hoyer, M.~H. van Kerkwijk, M.~Brett, A.~Haldane, J.~F. del R{\'{i}}o, M.~Wiebe, P.~Peterson, P.~G{\'{e}}rard-Marchant, K.~Sheppard, T.~Reddy, W.~Weckesser, H.~Abbasi, C.~Gohlke, and T.~E. Oliphant, {\em Nature}, 2020, {\bf 585}, 357--362.

\end{thebibliography}

\end{document}


\title{Supplementary material for General-order open-shell coupled-cluster method with partial-spin adaptation II: further formulations, simplifications, implementations, and numerical results}
\author{Cong Wang}
\email{congwang.webmail@gmail.com}
\affiliation{PO Box 26 Okemos, MI, 48805 (USA)}
\affiliation{The Pennsylvania State University; 401A Chemistry Building; University Park, PA 16802 (USA)}

\maketitle

\section{Naming conventions for equation files}

Naming conventions, e.g.,
\begin{align}
& \textrm{CCSDT\_os\_10\_r12t12\_eqns\_before\_permu-R.txt} \nonumber \\
& \textrm{CCSDT\_os\_10\_r12t12\_eqns\_permu-R.txt} \nonumber \\
& \textrm{CCSDT\_os\_10\_r12t12\_eqns\_permu-R\_pre.txt} \nonumber 
\end{align}

In the above file names,
\setlist[itemize]{leftmargin=2.5cm} 
\begin{itemize}
\item[CCSDT:] method level
\item[os:] open shell
\item[10:] sum of the orders of $\{R\} \{H\} \{T\}$ excitation operators. It is taken $E_{pq}^{rs}$ as 2 \cite{wang2018simple} and $E_{pq}^{rs} E_{ij}^{vw}$ as 4, similar to the closed-shell case \cite{wang2018simple}. For CCSD, CCSDT, and CCSDTQ levels, the maximum non-vanishing total orders for both closed-shell and partial-spin adaptation (PSA) are 8, 10, and 12, respectively. 
\item[r12t12:]  level of spin adaptation
\item[before\_permu-R:] no further usage of linear combinations of permutation structure of residual indices (similar to not adopt 5, -1, -1, -1,- 1, -1 in the closed-shell CCSDT in Figure 1 in Ref. \cite{wang2018simple}). 
\item[permu-R:]  used linear combinations of permutation structure of residual indices 
\item[permu-R\_pre:]   the prefactors of linear combinations of permutation structure of residual indices associated with the permu-R file. In each line, the first $N!$ where $N=2, 3$, and $4$ for double, triple, and quadruple excited residuals are to relate equations in permuting the first upper indices in the projection operator. For example, with $E_{ABC}^{IJV}$, the indices $IJV$ are permuted in terms of (012), (021), (102), (120), (201), (210) to compare the working equations.  For example, in CCSDT\_os\_10\_r12t12\_eqns\_before\_permu-R
\begin{align}
r[ABCVIJ] \, &+= \, 3/10 \, W[BCvw]  \, t[AvwVIJ] \nonumber \\
r[ABCVIJ] \, &+= \, -3/10 \, W[BCvw] \, t[AvwVJI] \nonumber
\end{align}
would be converted into 
\begin{align}
r[ABCVIJ] \, += \, 0.3 \, W[BCvw] \, t[AvwVIJ]  \nonumber
\end{align}
in CCSDT\_os\_10\_r12t12\_eqns\_permu-R.txt and 1.0, -1.0 in CCSDT\_os\_10

\_r12t12\_eqns\_permu-R\_pre.txt.

The subsequent entries in each line of the permu-R\_pre.txt file start from permuting lower indices, e.g., $E_{ACB}^{IJV}$. They are not used in the generating the permu-R file or computations of the working equations in the present work. These subsequent values are all zero. 
\end{itemize}

\section{Python protocol for tensor contractions}

The python file einsum\_time.py is a protocol to compare the performance of tensor contractions with small common dimensions in two approaches: (assuming Einstein summation rule that repeated indices as summation and '+=' symbol means a series of additions)
\begin{itemize}
    \item[(i)] a serial evaluations of contractions $C[i,k] += A_m[i,j] B_m[j,k]$
    \item[(ii)]  merging the matrices \textbf{A}$_m$ and \textbf{B}$_m$ into larger dimension arrays and perform tensor contraction once
\end{itemize}
, where $m$ presents the index of the series of contractions.
 
The opt\_einsum package \cite{daniel2018opt} with NumPy \cite{harris2020array} as the default backend has been used to avoid the timing of recomputing contraction paths. Comparisons with using NumPy and NumPy with optimize = True with different arrays are included.

To resemble the tensor contraction related to $r[\,] +=  W[aijv] \,\, t[avji]$ mentioned in the main text of the present work,  $n_i = n_k = 1000$, $n_j=3$, and 10 serial contractions ($n_m=10$) in approaches (i) was used in einsum\_time.py. The dimension of $n_i$ is set as assuming 10 inactive orbitals, the cubic power will be 1000. $n_j=3$ corresponds to three active orbitals. 10 serial contraction is estimated as 10 open-shell sectors (between 9 at the CCSDT and 14 at the CCSDTQ levels).

On a single-core with i7-5600U CPU (by export OMP\_NUM\_THREADS=1), the computational time for the approaches (i) and (ii) for the opt\_einsum example (averaged 50 times)  are $3.1\times10^{-2}$  and $3.1\times10^{-3}$ seconds, respectively. This result suggests a factor of 10 improvement.

\section*{References}
\bibliographystyle{pccp}
\bibliography{topic_os}